\documentclass[prd,twocolumn,nofootinbib,showpacs,preprintnumbers]{revtex4}
\usepackage{graphicx}
\usepackage{amsmath}
\usepackage{amssymb}

\newcommand{\sss}[1]{{\scriptscriptstyle{#1}}}
\newcommand{\be}{\begin{equation}}
\newcommand{\en}{\end{equation}}
\newcommand{\bea}{\begin{eqnarray}}
\newcommand{\ena}{\end{eqnarray}}

\newcommand{\mP}{m_{_{\mathrm{Pl}}}}

\newcommand{\setR}{\mathbb{R}}
\newcommand{\etal}{\textsl{et al.~}}
\newcommand{\uc}{\mathrm{c}}
\newcommand{\ub}{\mathrm{b}}
\newcommand{\us}{\mathrm{s}}
\newcommand{\uS}{\mathrm{S}}
\newcommand{\uT}{\mathrm{T}}
\newcommand{\uE}{\mathrm{E}}
\newcommand{\usssTE}{\sss{\uT \uE}}
\newcommand{\usssS}{\sss{\uS}}
\newcommand{\usssT}{\sss{\uT}}
\newcommand{\zero}{{_0}}
\newcommand{\one}{{_1}}
\newcommand{\two}{{_2}}

\newcommand{\lss}{\mathrm{lss}}
\newcommand{\dm}{\mathrm{dm}}
\newcommand{\ess}{{_\mathrm{S}}}
\newcommand{\ti}{{_\mathrm{T}}}
\newcommand{\muS}{\mu_\ess}
\newcommand{\muT}{\mu_\ti}
\newcommand{\muST}{\mu_{\ess,\ti}}
\newcommand{\omegaT}{\omega_\ti}
\newcommand{\omegaS}{\omega_\ess}
\newcommand{\omegaST}{\omega_{\ess, \ti}}

\newcommand{\calH}{\mathcal{H}}
\newcommand{\Mc}{M_\uc}
\newcommand{\OmegaL}{\Omega_\Lambda}
\newcommand{\OmegaCDM}{\Omega_\dm}
\newcommand{\OmegaB}{\Omega_\ub}

\newcommand{\Ps}{P_\mathrm{scalar}}
\newcommand{\ns}{n_\us}
\newcommand{\epsilonT}{\epsilon_\two}
\newcommand{\CAMB}{\texttt{CAMB} }
\newcommand{\COSMOMC}{\texttt{COSMOMC} }
\def\setC{\mathbb{C}}

\def\setR{\mathbb{R}}

\newcommand{\GReCO}{${\cal G}\setR\varepsilon\setC{\cal O}$}

\begin{document}

\title{Superimposed Oscillations in the WMAP Data?}

\author{J\'er\^ome Martin} 
\email{jmartin@iap.fr}
\affiliation{Institut d'Astrophysique de Paris, \GReCO --FRE 2435, 
98bis boulevard Arago, 75014 Paris, France} 

\author{Christophe Ringeval}
\email{christophe.ringeval@physics.unige.ch}
\affiliation{D\'epartement de Physique Th\'eorique, Universit\'e 
de Gen\`eve, 24 quai Ernest Ansermet, 1211 Gen\`eve 4, Switzerland}

\date{\today} 

\begin{abstract}
The possibility that the cosmic variance outliers present in the
recently released WMAP multipole moments are due to oscillations in
the primordial power spectrum is investigated. Since the most
important contribution to the WMAP likelihood originates from the
outliers at relatively small angular scale (around the first Doppler
peak), special attention is paid to these in contrast with previous
studies on the subject which have concentrated on the large scales
outliers only (i.e. the quadrupole and octupole). As a physically
motivated example, the case where the oscillations are of
trans-Planckian origin is considered.  It is shown that the presence
of the oscillations causes an important drop in the WMAP $\chi ^2 $ of
about fifteen. The F-test reveals that such a drop has a probability
less than $0.06\% $ to occur by chance and can therefore be considered
as statistically significant.

\end{abstract}

\maketitle

\section{Introduction}
The recently released WMAP data~\cite{wmap} have confirmed the
standard paradigm of adiabatic scale invariant primordial
fluctuations~\cite{hinshaw,verde}. This paradigm can be justified in
the framework of inflation and can explain the most important
cosmological observations~\cite{peiris,saminf}. This remarkable
success has led the cosmologists to take an interest in more subtle
features of the WMAP multipole moments. In particular, recently, many
studies have been devoted to the so-called cosmic variance outliers,
i.e. points which lie outside the one sigma cosmic variance
error~\cite{outEf, outCon, outCline, outFeng,
TMB,alain,huang03,Ef}. These outliers have been considered as
intriguing since the probability for their presence would be very
small~\cite{spergel}. So far, all the studies have concentrated on the
seeming lack of power at large scales, i.e. on the quadrupole and the
octupole outliers. In the literature, two possibilities have been
envisaged. In Ref.~\cite{Ef}, it has been argued that the outliers are
not a problem at all, the crucial point being the way the probability
of their presence is estimated. In
Refs.~\cite{outCon,outCline,outFeng,TMB}, it has been envisaged that
the outliers could be a signature of new physics even if it has also
been recognized in these articles that the cosmic variance could be
responsible for their presence. In particular, in Ref.~\cite{outCon},
it has been proposed that the inflationary scale invariant initial
power spectrum could be modified by some new physics such that a sharp
cut-off at large scales appears while it remains unchanged
elsewhere. It has been shown that this can cause a decrease of the
$\chi ^2$ of order $\Delta \chi ^2\simeq 2-4$ for one additional free
parameter given by the scale of the cut-off.

\par

However, as revealed by the Fig.~4 of Ref.~\cite{spergel}, the main
contribution to the WMAP $\chi ^2$ does not come from the large scales
but rather from scales which correspond to the first and second
Doppler peaks (more precisely, according to Ref.~\cite{spergel}, the
three main contributions come from the angular scales $\ell \simeq
120, 200$ and $340$). In other words, if the presence of outliers is
taken seriously into account, modifications of the standard power
spectrum seem to be required on a different range of scales and, in
any case, not only at very large scales. In addition, the small scale
outliers can be above or below the theoretical error bar and,
therefore, the required modifications do not seem to have the form of
a systematic lack or excess of power. This naturally leads to the idea
that the power spectrum could possess superimposed oscillations. The
aim of this article is to study whether this idea has any statistical
support. Of course, a physical justification for the presence of
oscillations in the power spectrum is needed. Interestingly enough,
such a justification exists precisely in the context of the theory of
inflation~\cite{wang,MB1, BM1} and we now discuss this question in
more details.

\par

One of the main advantage of the inflationary scenario is that it
permits to fix sensible initial conditions. Because the Hubble radius
was constant during inflation, the wavelength of a mode of
astrophysical interest today was much smaller than the Hubble scale at
the beginning of inflation whereas, without a phase of inflation, the
same mode would have been a super-Hubble mode. Contrary to the
super-Hubble case, the vacuum is defined without ambiguity in the
sub-Hubble regime. This state is the starting point of the subsequent
cosmological perturbations evolution. This leads to a nearly
scale-invariant spectrum for density perturbations, a prediction which
is now confirmed to a high level of
accuracy~\cite{wmap,hinshaw,verde,peiris,saminf}.

\par

However, this remarkable success carries in itself a potential
problem. In a typical model of inflation, the modes are initially not
only sub-Hubble but also sub-Planckian, that is to say their
wavelength is smaller than the Planck length
$\mP^{-1}$~\cite{MB1,BM1}. In this regime, the physical principles
which underlie the calculations of the power spectrum are likely not
to be valid anymore. This problem is specific to the perturbative
approach and does not affect the background model. Indeed, since the
energy density of the inflaton field is nearly constant during
inflation, we face in fact a situation where the wavelength of the
Fourier modes is smaller than the Planck length whereas the energy
density which drives inflation can still be well below the Planck
energy density. The problem described above mainly concerns the modes
of cosmological (still in the linear regime) interest today which are,
in the inflationary paradigm, a pure relic of the trans-Planckian
regime. It should be added that the scale at which the new physics
shows up is not necessarily the Planck length but could be everywhere
between this scale and the Hubble radius. In this article we denote
the new energy scale $\Mc$ and assume that $\sigma_\zero
\equiv H/\Mc$ is a free parameter ($H$ denotes the Hubble
parameter during inflation). If the physics is different beyond the
scale $\Mc$, this should leave some imprints on the spectrum
of inflationary cosmological perturbations and therefore modify the
Cosmic Microwave Background (CMB) anisotropies~\cite{wang,MB1,BM1}.

\par

The next question is how to calculate these modifications. Roughly
speaking, the calculation of the power spectrum of the fluctuations
reduces to the calculation of the evolution of a free quantum scalar
field in a time-dependent background. Various methods have been used
to model the physics beyond $\Mc$ like changing the free
dispersion relation~\cite{MB1,BM1,N,LLMU}, using stringy uncertainty
relations~\cite{Kempf,EGKS,EGKS2, KN,BH,HS} or noncommutative
geometry~\cite{LMMP}. Yet another approach has been to assume
that the Fourier modes are ``created'' when their wavelength equals
the critical scale $\Mc^{-1}$~\cite{D,EGKS3,ACT,AL,BM03}. A
generic prediction, first made in Refs.~\cite{MB1,BM1}, which appears
to be independent of the settings used to model the new physics, is
the presence of oscillations in the power spectrum (of course, the
detailed properties of these oscillations do depend on the model
utilized). This can be understood easily. In the standard calculation,
initially, the scalar field is just given by an in-going wave. If the
evolution proceeds according to the WKB approximation~\cite{MSwkb}, at
Hubble scale exit the scalar field will just differ by a phase which
will drop out when the power spectrum is calculated. On the other
hand, if the WKB approximation is violated at some time before Hubble
scale exit or if the modes are created at a fixed length in some
$\alpha $-vacuum, then the scalar field will be a combination of
on-going and out-going waves. This combination gives the oscillations
in the power spectrum. To put it differently, one can say that the
existence of a preferred scale plus the standard inflationary theory
generically imply a power spectrum given by a nearly scale invariant
component plus superimposed oscillations.

\par

The aim of this article is to use the previous generic trans-Planckian
prediction as a tool to analyze whether the WMAP multipole moments
exhibit oscillations originating from oscillations in the primordial
spectrum. It should be stressed again that, in this work, the
trans-Planckian effects are only considered as an example of models
where oscillations could show up. Indeed, the analysis presented in
this article would still be useful for any alternative model which
predicts a similar oscillatory pattern.  

This paper is organized as follows. In the next section, we briefly
recall the standard calculation of the inflationary spectrum when the
trans-Planckian effects are taken into account. In
Sect.~\ref{sect:cmb}, we compare the theoretical predictions with the
WMAP data. Finally, we discuss the data analysis and give our
conclusions in Sect.~\ref{sect:end}.

\section{The trans-Planckian power spectra}
\label{sect:ps}

In this article, we assume that inflation of the spatially flat FLRW
spacetime is driven by a single scalar field $\phi _\zero(\eta
)$. Scalar perturbations of the geometry can be characterized by the
gauge-invariant Bardeen potentials $\Phi $ and fluctuations in the
scalar field are characterized by the gauge-invariant quantity
$\delta\phi $. If $\phi _\zero'\neq 0$ (a prime denotes derivative
with respect to conformal time $\eta$), then everything can be reduced
to the study of a single gauge-invariant variable (the so-called
Mukhanov-Sasaki variable) defined by \cite{MFB} $v\equiv a(\delta
\phi+ \phi _\zero'\Phi/{\calH})$, where $\calH=aH$ is the conformal
Hubble parameter. In fact, it turns out to be more convenient to work
with the variable $\muS$ defined by $\muS\equiv
-\sqrt{2\kappa }v$, where $\kappa \equiv 8\pi/\mP^2$. Density
perturbations are often characterized by the so-called conserved
quantity $\zeta $~\cite{MS} defined by $\zeta \equiv (2/3)({\cal
H}^{-1}\Phi '+\Phi )/(1+w )+\Phi $, where $w $ is the equation of
state parameter. The quantity $\muS$ is related to $\zeta$ by
$\muS=-2a\sqrt{\gamma }\zeta $, where $\gamma
=1-{\calH}'/{\calH}^2$. On the other hand, the primordial
gravitational waves are described by the transverse and traceless
tensor $h_{ij}$, the Fourier transform of which is $\muT$ (up to a
polarization tensor). Both types of perturbations obey the same type
of equation of motion, namely the equation of a parametric
oscillator~\cite{MS}
\begin{equation}
\label{eq:evol}
\muST''+\omegaST^2(k,\eta ) \muST=0 ,
\end{equation}
with $\omegaS^2=k^2 -(a\sqrt{\gamma })''/(a\sqrt{\gamma })$,
$\omegaT^2=k^2 -a''/a$, and $k$ is the wavenumber of a given Fourier
mode.
Finally, the quantities of interest for computing the CMB
anisotropies are the power spectra which read~\cite{MS} 
\begin{equation}
\label{spec}
k^3P_{\zeta }(k)=\frac{k^3}{8\pi ^2}\biggl\vert
\frac{\muS}{a\sqrt{\gamma }}\biggr\vert ^2 , \quad
k^3P_h(k)=\frac{2k^3}{\pi ^2}\biggl \vert \frac{\muT}{a}
\biggr \vert ^2 .
\end{equation}
In order to compute $k^3P_{\zeta }(k)$ and $k^3P_h(k)$, one must
integrate the equation of motion (\ref{eq:evol}) and specify what the
initial conditions are.  The integration of the equations of motion is
possible for a large class of inflationary models provided they
satisfy the slow-roll conditions~\cite{SL}.  The slow-roll
approximation is controlled by a set of parameters given by $\epsilon
\equiv \gamma$, $\delta \equiv \phi''_\zero/(\calH \phi'_\zero)-1$ and
$\xi \equiv (\epsilon'-\delta')/\calH$. The slow-roll conditions are
satisfied if $\epsilon$ and $\delta$ are much smaller than one and if
$\xi = {\cal O}(\epsilon^2,\delta^2,\epsilon\delta)$. At first order,
the parameters $\epsilon $ and $\delta $ can be considered as
constant.

\par

Here, the initial conditions are fixed under the assumption that the
Fourier modes never penetrate the trans-Planckian region. In other
words, a Fourier mode is supposed to ``appear'' when its wavelength
becomes equal to a new fundamental characteristic scale
$\ell_{\mathrm{c}}$. The time $\eta _k$ of mode ``appearance'' with
comoving wavenumber $k$, can be computed from the condition
\begin{equation}
\lambda (\eta _k)=\frac{2\pi }{k}a(\eta _k)=\ell _{\mathrm{c}}
\equiv \frac{2\pi }{\Mc} ,
\end{equation}
which implies that $\eta _k$ is a function of $k$. This has to be
compared with the standard inflationary calculations where the initial
time is taken to be $\eta _k=-\infty $ for any Fourier mode $k$ and
where, in a certain sense, the initial time does not depend on $k$
(see, however, Ref.~\cite{chung}). In the framework of trans-Planckian
inflation, a crucial question is in which state the Fourier mode is
created at the time $\eta _k$. Here, we consider the most general
conditions, also called a truncated
$\alpha$\---vacuum~\cite{BMa,EL,KKLSS,GL,CHM,CR}
\begin{eqnarray}
\label{ci1}
\muST(\eta _k) &=& \mp
\frac{c_k+d_k}{\sqrt{2\omega _{_{\rm S,T}}(\eta _k)}}
\frac{4\sqrt{\pi }}{m_{_{\rm Pl}}} , 
\\
\label{ci2}
\muST'(\eta _k) &=&
\pm i\sqrt{\frac{\omega _{_{\rm S,T}}(\eta _k)}{2}}
\frac{4\sqrt{\pi }(c_k-d_k)}{m_{_{\rm Pl}}} .
\end{eqnarray}
The coefficients $c_k$ and $d_k$ are {\it a priori} two arbitrary
complex numbers satisfying the condition $\vert c_k\vert ^2-\vert
d_k\vert ^2=1$. Since there are two energy scales in the problem,
namely the Hubble parameter $H$ during inflation and the new scale
$\Mc$, the final result will be expressed in terms of the ratio
$\sigma_\zero\equiv H/\Mc$, which is a small parameter. As a result,
we typically expect that $c_k=1+y\sigma_\zero+\cdots $ and
$d_k=x\sigma_\zero+\cdots $ since the initial conditions are expressed
at time $\eta _k$~\cite{AL}.
The parameters $x$ and $y$ are considered
as free parameters that are not fixed by any existing well-established
theories (see also Ref.~\cite{AL}) except, of course, that they should
be such that the relation $\vert c_k\vert ^2-\vert d_k\vert ^2=1$ is
satisfied. One easily shows that this implies $y+y^*=0$ at leading
order in $\sigma _\zero $. Expanding everything in terms of
$\sigma_\zero$, one arrives at~\cite{BM03}
\vspace{-1cm}
\begin{widetext}
\begin{eqnarray}
\label{pssrs2}
k^3P_{\zeta } &=&\frac{H^2}{\pi \epsilon m_{_{\rm Pl}}^2}
\biggl\{1-2(C+1)\epsilon -2C(\epsilon -\delta )-2(2\epsilon -\delta )
\ln \frac{k}{k_*}-2\vert x\vert \sigma _\zero\biggl[
1-2(C+1)\epsilon -2C(\epsilon -\delta )
-  2(2\epsilon -\delta )\ln \frac{k}{k_*}\biggr]\nonumber \\
& \times & \cos \biggl[\frac{2}{\sigma _\zero}
\biggl(1+\epsilon +\epsilon \ln \frac{k}{a_\zero \Mc}\biggr)
+\varphi \biggr]
-2\vert x\vert \sigma _\zero\pi (2\epsilon -\delta )
\sin \biggl[\frac{2}{\sigma _\zero}
\biggl(1+\epsilon +\epsilon \ln \frac{k}{a_\zero \Mc}\biggr)
+\varphi \biggr]
\biggr\},\\
\label{pst}
k^3P_h &=& \frac{16 H^2}{\pi m_{_{\rm Pl}}^2}
\biggl\{1-2(C+1)\epsilon -2\epsilon \ln \frac{k}{k_*}
-2\vert x\vert \sigma _\zero\biggl[1-2(C+1)\epsilon -2\epsilon 
\ln \frac{k}{k_*}\biggr]
\cos \biggl[\frac{2}{\sigma _\zero}
\biggl(1+\epsilon +\epsilon \ln 
\frac{k}{a_\zero \Mc}\biggr)+\varphi \biggr]
\nonumber \\
& - & 2\vert x\vert \sigma _\zero\pi\epsilon
\sin \biggl[\frac{2}{\sigma _\zero}
\biggl(1+\epsilon +\epsilon \ln \frac{k}{a_\zero \Mc}\biggr)
+\varphi \biggr]
\biggr\}\, ,
\end{eqnarray}
\end{widetext}
where $\varphi $ is the argument of the complex number $x$, i.e
$x\equiv \vert x\vert {\rm e}^{i\varphi }$. The constant $C$ is given
by $C\equiv \gamma _{_{\rm E}}+\ln 2-2$, $\gamma _{_{\rm E}}$ being
the Euler constant. The scales $k_*$ is the pivot
scale~\cite{MS2,MRS}. The parameter $\sigma _\zero$ and the scale
factor $a_\zero$ are evaluated at the time $\eta _\zero$ during
inflation which is {\it a priori} arbitrary but, and this is the
important point, does not depend on $k$. In the following we will
choose this time such that $k_*/a_\zero=\Mc$. One sees that we obtain
a scale-invariant spectrum plus logarithmic corrections in $k$ the
amplitude of which is determined by the slow-roll parameters. In
addition, superimposed oscillations coming from the trans-Planckian
initial conditions of Eqs.~(\ref{ci1}) and (\ref{ci2}) appear. The
magnitude of the trans-Planckian corrections are linear in the
parameter $\sigma _\zero$ and their amplitude is given by $\vert
x\vert \sigma _\zero$, a result in agreement with Ref.~\cite{AL}. Let
us also remark that the parameter $y$ does not appear at the leading
order considered in Eqs.~(\ref{pssrs2}) and (\ref{pst}). The
wavelength of the oscillations can be expressed as
\begin{equation}
\label{wl}
\frac{\Delta k}{k}=\frac{\sigma _\zero\pi }{\epsilon}\, ,
\end{equation}
in agreement with Ref.~\cite{D2}. The presence of the factor $\vert
x\vert $ in the amplitude of the trans-Planckian corrections plays an
important role in what follows. As stressed in Ref.~\cite{AL},
Refs.~\cite{D,D2,steen} assume that $\vert x\vert =1$. This has the
consequence that high frequency waves have necessarily a small
amplitude [see Eq.~(\ref{wl})]. The factor $\vert x\vert$ allows to
break this degeneracy. As a result the amplitude and the wavelength of
the corrections become two independent quantities. The data analysis
can then proceed in a larger parameter space. This explains why we can
obtain results different from those derived in Ref.~\cite{steen}
without being in contradiction with that article. However, let us also
remark that a coefficient $\vert x\vert \neq 1$ implies that the
corresponding initial state cannot belong to the class of vacua often
considered in previous works as, for instance, the instantaneous
Minkowski vacuum state or the initial state considered in
Ref.~\cite{D}. In this case, and as it will be discussed below, we
face the issue of back-reaction~\cite{BMa,EL,KKLSS,GL}. Finally, let
us also notice that the power spectra depend on both parameters
$\epsilon $ and $\delta $ and not only on the parameter $\epsilon $,
contrary to some results recently obtained in the literature.

\par

The initial power spectra of Eqs.~(\ref{pssrs2}) and (\ref{pst}) are
related to the CMB anisotropy through the multipole moments which are
in turn defined through the two-point correlation function of the
temperature fluctuations. Explicitly, we have
\begin{equation}
\biggl \langle \frac{\delta T}{T}({\bf e}_1)\frac{\delta T}{T}({\bf e}_2)
\biggr \rangle =\sum _{\ell =2}^{+\infty }\frac{2\ell +1}{4\pi }C_{\ell }
P_{\ell }(\cos \theta )\, ,
\end{equation}
where $\theta $ is the angle between the two directions ${\bf e}_1$
and ${\bf e}_2$. The WMAP satellite has also measured the polarization
and therefore we will also be interested in the ``TE'' cross
polarization multipole moments $C_{\ell
}^{\usssTE}$~\cite{wmapte}. The (temperature) multipole moments are
the sum of the scalar contribution and of the tensor contribution,
$C_{\ell }= C_{\ell }^{\usssS}+C_{\ell }^{\usssT}$. From
Eqs.~(\ref{pssrs2}) and (\ref{pst}), we see that in the context of
slow-roll inflation, with or without trans-Planckian corrections, the
contribution of gravitational waves cannot be put to zero
arbitrarily. Indeed, the ratio of the two types of perturbations is
predicted by the form of the power spectra and reads
$k^3P_h/k^3P_{\zeta }=16\epsilon (1+\cdots )$, where the dots stand
for the trans-Planckian corrections. The fact that the trans-Planckian
physics modifies the consistency check of inflation was first noticed
in Ref.~\cite{HK}. Finally, we notice that, since we work at first
order in the slow-roll parameters, there is no running of the spectral
indices.

\par

The derivation of the trans-Planckian corrections in the power spectra
(\ref{pssrs2}) and (\ref{pst}) assumes that the backreaction effects
are not too important. For consistency, the energy density of the
perturbations must be smaller or equal than that of the inflationary
background. This leads to the condition $\vert x\vert \le \sqrt{3\pi}
\mP/\Mc$, an estimate which is in agreement with that derived in
Ref.~\cite{AL}. In order to put numerics on the above constraint, we
can use the large scale approximation of the multipole moments
\begin{equation}
\label{eq:largescale}
C_{\ell }^\usssS \simeq \frac{4\pi }{25}\int _0^{+\infty}j_{\ell
}^2(kr_{\rm lss}) k^3P_{\zeta }\frac{{\rm d}k}{k}\, ,
\end{equation}
where $j_{\ell }$ is a spherical Bessel function of order $\ell $ and
$r_{\rm lss}$ is the distance to the last scattering surface.  This
approximation is valid for $\ell \lesssim 10$ and also requires a
vanishing cosmological constant with no integrated Sachs-Wolfe effect.
Neglecting the logarithmic corrections in the power spectrum gives
$\ell (\ell +1)C_{\ell }^{_{\rm S}} \simeq 2H^2/(25\epsilon m_{_{\rm
Pl}}^2)$. On the other hand, since the COBE-WMAP normalization is
$6T^2C_2^{_{\rm S}}\times 10^{12}/(2\pi )\simeq 10^3$, where $T\simeq
2.72\mbox{K}$, this implies that $H\simeq m_{_{\rm Pl}}\sqrt{\epsilon
}10^{-4}$. Finally, using the fact that $\Mc=H/\sigma _0$, one arrives
at
\begin{equation}
\label{eq:brvalid}
\vert x\vert \sigma _0\le 10^4 \times 
\frac{\sigma _0^2}{\sqrt{\epsilon }} ,
\end{equation}
where, in order to derive an order of magnitude estimate, the
unimportant factors of order one have been neglected. It is important
to emphasize that the above constraint is only a sufficient condition,
but by no means, unless proven otherwise, a necessary condition for
the validity of the power spectra calculations.

\section{Comparison with the CMB data}
\label{sect:cmb}

In this section, the compatibility of the trans-Planckian models
described in the previous section with the WMAP data is studied. More
precisely, we now use the trans-Planckian power spectra derived
previously to test the idea that the cosmic variance outliers could be
due to oscillations in the WMAP data. In order to compute the
multipole moments $C_{\ell }$ and $C_{\ell }^{\usssTE}$, we have used
a modified version of the \CAMB code~\cite{camb} based on its
slow-roll inflation module~\cite{sam}. Following the analysis of
Ref.~\cite{outCline}, the parameters space that we consider is $h$,
$\Omega _{\rm b}$, $\Omega _{\rm dm}$, $z_{\rm re}$ (or $\tau $),
$\epsilon $, $\delta $, $\Ps$ and the trans-Planckian parameters
$\sigma _\zero$, $x$. The parameter $h$ is the dimensionless Hubble
constant, $\Omega _{\rm b}$ the amount of baryons, $\Omega _{\rm dm}$
the amount of dark matter and $z_{\rm re}$ the redshift of
reionization. $\Ps$ is the normalization of the scalar power spectrum,
and as mentioned above, it also fixes the normalization of the
gravitational waves. We have restricted ourselves to flat models,
i.e. the cosmological constant is given by $\Omega _{\Lambda}=
1-\Omega _{\rm dm}-\Omega _{\rm b}$. The choice $\varphi =0$ has been
made since we have checked that the value of $\varphi$ has no real
influence for our purpose. Therefore, we deal with a $9$-dimensional
parameter space. In order to estimate the parameters, we have used
Monte-Carlo methods implemented in the \COSMOMC code~\cite{cosmomc}
with our modified \CAMB version, together with the likelihood code
developed by the WMAP team~\cite{verde,hinshaw}.

\par

We have first checked that, without trans-Planckian effects, i.e. by
setting $|x|\sigma_\zero=0$ in the expressions of the scalar and
tensor power spectra, the standard results are recovered. The best
WMAP fit given in Ref.~\cite{spergel} is well reproduced, as well as
the values for the slow-roll parameters derived in
Refs.~\cite{peiris,saminf,barger}. Concerning the likelihood
computation, by setting the input parameters to the best fit values
given in Ref.~\cite{spergel}, we obtain $\chi ^2\simeq 1430.92$ for
$1342$ degrees of freedom, also in good agreement with
Ref.~\cite{spergel} (see the first line in Table~\ref{tbl:bestfit}).
\begin{table*}
\begin{tabular}{|r|c|c|c|c|c|c|c|c|c|c|c|c|}
\hline $ |x|\sigma_\zero$ & $\sigma_\zero$ & $h$ & $\OmegaB h^2$ &
$\OmegaCDM h^2 $ & $\OmegaL$ & $\tau$& $\Ps$ &
$\epsilon_\one=\epsilon$ & $\delta$ &
$\epsilonT=2(\epsilon-\delta)$ &$\ns\equiv1-4\epsilon + 2\delta$
& $\chi^2/\mathrm{d.o.f.}$ \\ \hline $0$ & $-$ & $0.68$ & $0.023$ &
$0.127$ & $0.66$ & $0.11$ & $24.5 \times 10^{-10}$ & $0.001$ &
$-0.013$ & $0.028$ & $0.97$ & $1430.92/1342$\\ $0.29$ & $1.6 \times
10^{-4}$ & $0.79$ & $0.025$ & $0.103$ & $0.79$ & $0.21$ & $25.2\times
10^{-10}$ & $0.020$ & $0.061$ & $-0.083$ & $1.04$ & $1415.38/1340$\\
\hline
\end{tabular}
\caption{Best fit parameters from the WMAP data for the standard model
of inflation compared to the best fit parameters obtained for the
trans-Planckian model. The pivot scale and the time $\eta _\zero$ (see
the text) are chosen such that $k_*/a_\zero =\Mc
=0.01\mbox{Mpc}^{-1}$. The quantities $\epsilon _1$ and $\epsilon _2$
are two different slow-roll parameters, defined in Ref.~\cite{STG},
and used in Ref.~\cite{saminf} to express the observational
constraints obtained from the WMAP data. These parameters are linearly
related with $\epsilon $ and $\delta $.}
\label{tbl:bestfit}
\end{table*}

\par

In a second step, we have taken into account the oscillations present
in the primordial power spectrum. Roughly speaking, the multipole
moments $C_{\ell }$ are given by the convolution of a spherical Bessel
function with the initial power spectrum [see
Eq.~(\ref{eq:largescale})]. It is therefore not obvious that the
oscillations in the primordial spectra will be transferred into
oscillations in the multipole moments. As shown in Ref.~\cite{Hu03},
fine structures in the initial power spectra are strongly broadened
and damped by the transfer function. However, the oscillations
considered here are not localized at a given momentum scale but are
rather spread all over the $k$--space. One may therefore expect
constructive interferences to appear at some angular scale $\ell $
even in presence of the damping mentioned above. From the analytical
point of view, the integral in Eq.~(\ref{eq:largescale}) can be
evaluated exactly for the case $2\epsilon=\delta $ (i.e. $\ns=1$) and
this will allows us to gain some intuition with regards to the
behavior of the trans-Planckian multipole moments. They can be
expressed as
\begin{widetext}
\begin{equation}
\label{multi}
\begin{aligned}
\ell (\ell +1)C_{\ell }^{\usssS} \simeq \frac{2H^2}{25\epsilon \mP^2}
(1-2\epsilon ) &\left[1- \vert x\vert \sigma_\zero \sqrt{\pi} \ell (\ell
+1)\, \Re \negthickspace \left( \dfrac{\exp{\left\{i \dfrac{2}{\sigma_\zero}[1+\epsilon
- \epsilon \ln (a_\zero \Mc r_\lss) ] + i \varphi
\right\}}}{\left(\ell - i\dfrac{\epsilon}{\sigma_\zero}\right)
\left(\ell + 1 - i\dfrac{\epsilon}{\sigma_\zero}\right)} \right. \right. \\
& \times \left. \left.
\dfrac{\Gamma \left(1-i\dfrac{\epsilon}{\sigma_\zero}\right) \Gamma
\left(\ell +i \dfrac{\epsilon}{\sigma\zero}\right)}{\Gamma \left(
\dfrac{3}{2} - i\dfrac{\epsilon}{\sigma_\zero} \right) \Gamma
\left(\ell -i \dfrac{\epsilon}{\sigma_\zero} \right)}\right)\right],
\end{aligned}
\end{equation}
\end{widetext}
where $\Gamma$ stands for the Euler's integral of second
kind~\cite{GR}, and with $\ell \lesssim 10$. Although this equation is
not specially illuminating, we can already conclude that, due to the
presence of the Euler functions with complex arguments, the
oscillations are indeed passed to the multipole moments. This is
already an important conclusion since this means that the oscillations
in the primordial power spectrum are not killed by the convolution. In
the limit $\epsilon/\sigma_\zero \gg \ell$, the above formula
simplifies a lot and reads
\begin{equation}
\label{eq:largescalestpl}
\begin{aligned}
\ell(\ell & +1) C_\ell^{\usssS}  \simeq \dfrac{2 H^2}{25 \epsilon \mP^2}
(1-2\epsilon) \Bigg\{ 1 + \sqrt{\pi} \dfrac{|x|\sigma_\zero
\ell(\ell+1)}{(\epsilon/\sigma_\zero)^{5/2}} \\ & \times \cos\left[\pi
\ell + \dfrac{2}{\sigma_\zero}\left(1 + \epsilon\ln
\dfrac{\epsilon/\sigma_\zero}{a_\zero \Mc r_\lss} \right)
+ \varphi - \dfrac{\pi}{4}\right] \Bigg\}.
\end{aligned}
\end{equation}
One sees that, at small $\ell $, these oscillations are
damped by a factor of $(\epsilon/\sigma_\zero)^{5/2}$, for a given
primordial amplitude $|x|\sigma_\zero$. Note that at larger $\ell $,
this effect may be compensated by the factor $\ell (\ell
+1)$. Therefore, in practice, we expect the oscillations to appear
only at relatively small scales and to be absent at large angular
scales. Of course, at large $\ell $, or for not too small values of
$\sigma_\zero$, the above equation quickly becomes invalid and an
accurate estimation can be made only with the help of numerical
calculations.

\par

\begin{figure}
\includegraphics[angle=270,width=9.6cm]{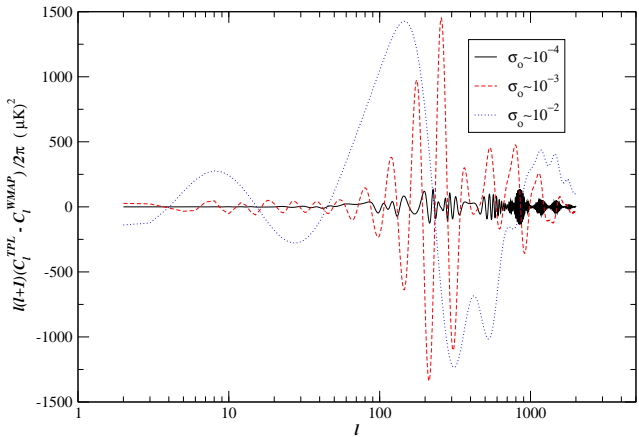}
\caption{Difference between the trans-Planckian and the standard
inflationary multipole moments, plotted for three values of
$\sigma_\zero$ and for an amplitude of $\vert x\vert \sigma
_\zero=0.3$. A strong damping of the oscillations is observed at large
angular scales.}
\label{fig:deltaCl}
\end{figure}

\begin{figure*}
\includegraphics[angle=0,width=18.8cm]{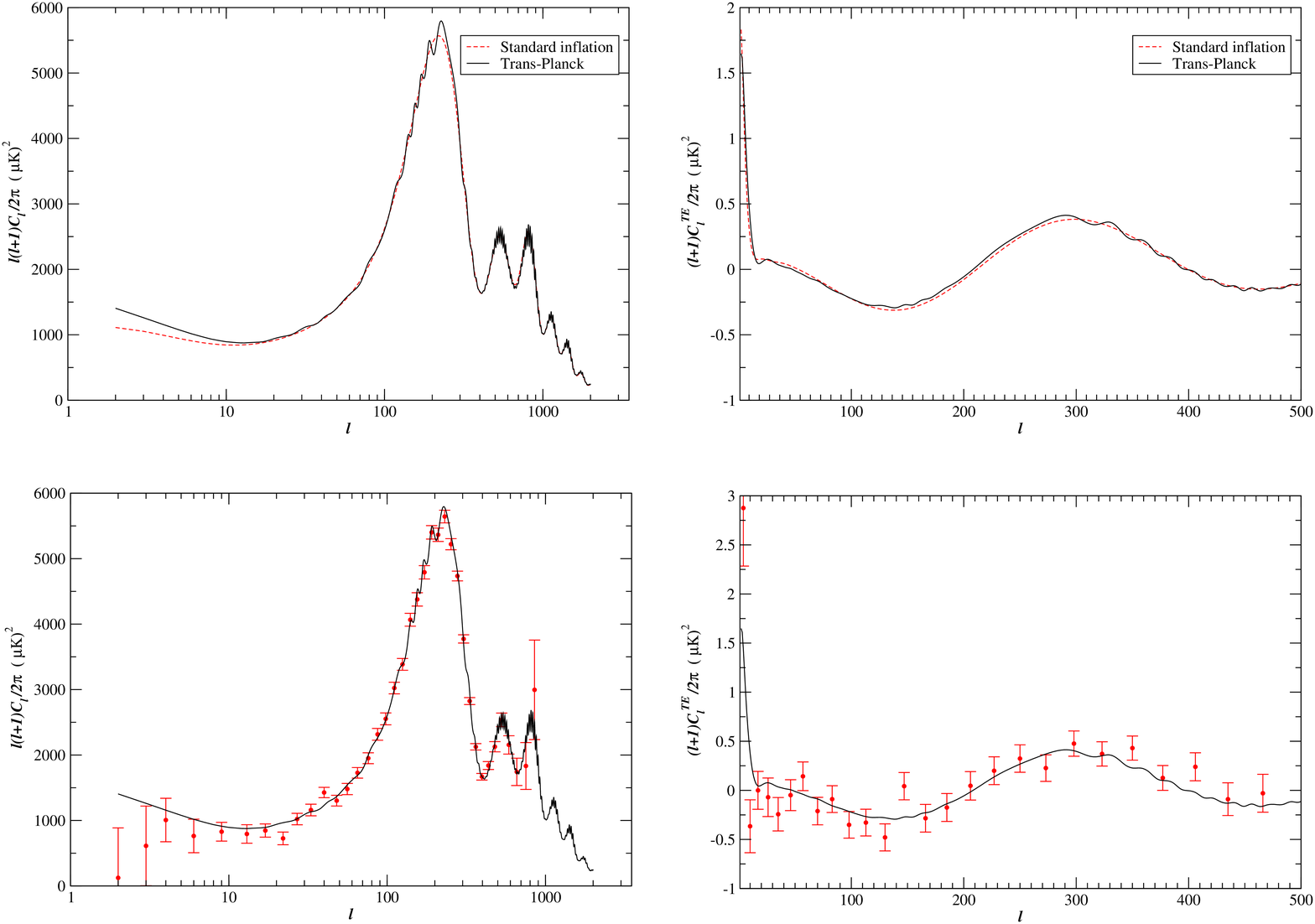}
\caption{Top left and right panels: the best fit obtained for
trans-Planckian angular TT and TE power spectra compared to the best
fit standard inflationary power spectra. Bottom left and right panels:
the same best fit trans-Planckian TT and TE power spectra but this
time compared to the binned WMAP data~\cite{wmap}.}
\label{fig:cls}
\end{figure*}

These computations are not trivial and require the \CAMB accuracy to
be increased in order to correctly transfer the superimposed
oscillations in the power spectra to the multipole moments. Indeed,
the \CAMB default computation target accuracy of $1\%$ for the scalars
is only valid for non-extreme models. Since we precisely consider fine
structures in the power spectra, the corresponding models do not enter
into this class and modifications in the default values of the \CAMB
code parameters are therefore required for small values of $\sigma
_\zero $. Without these modifications (i.e. with the default
accuracy), random structures in the multipole moments were found to
appear at large scales for models with $\sigma_\zero<10^{-2}$. The
modified accuracy of the code has been chosen such that it allows to
treat models with sufficiently small values of $\sigma _\zero$ (in
practice, up to $\sigma_\zero = 10^{-4}$) and to avoid prohibitive
computation time. Then, the strong damping of the high frequency
oscillations is recovered at large scale. In Fig.~\ref{fig:deltaCl},
the difference between the trans-Planckian and inflationary angular
power spectra, $C_\ell^{\sss{\mathrm{TPL}}} -
C_\ell^{\sss{\mathrm{WMAP}}}$, has been plotted for three different
values of $\sigma_\zero$. As mentioned above, the previous qualitative
considerations are recovered: the smaller $\sigma_\zero$ and $\ell$,
the more damped the oscillations in the angular power spectrum. For
values of $\sigma_\zero < 10^{-3}$, the remaining oscillations are
even washed out at small $\ell $ and only show up at the rise of the
first acoustic peak. We also notice a similar behavior for $C_{\ell
}^{\usssTE}$ (see Fig.~\ref{fig:cls}).

\begin{figure}
\includegraphics[angle=270,width=9.6cm]{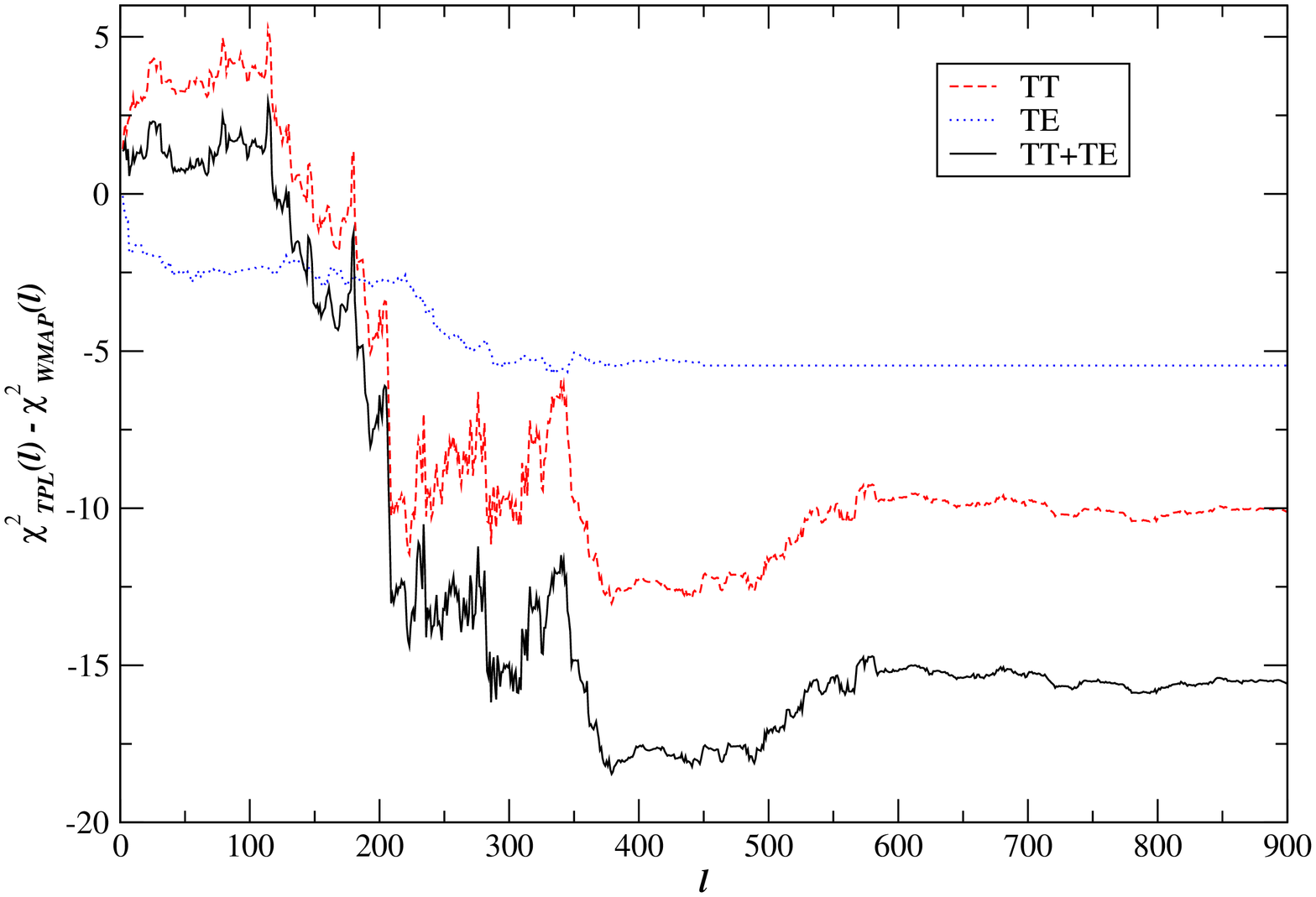}
\caption{Cumulative residual $\Delta \chi^2(\ell)$ between the
trans-Planckian fit and the WMAP best fit. The contributions stemming
from the temperature and the polarization angular power spectra are
also plotted. The final $\Delta\chi^2$ improvement is of order $15$.}
\label{fig:cumulchi2}
\end{figure}

Having checked that the numerical calculations are well under control,
one can now move to parameters estimation. The main result of this
article is that, with the oscillations taken into account, it is
possible to decrease the $\chi ^2$ significantly. The best fit that
has been found by \COSMOMC and the result of its Markov chains
exploration of the parameter space is summarized in
Table~\ref{tbl:bestfit}. It leads to $\chi^2\simeq 1415.38$ for $1340$
degrees of freedom, i.e. $\Delta \chi ^2\simeq 15$ compared to WMAP
one. The corresponding multipole moments $C_{\ell }$ and $C_{\ell
}^{\usssTE}$ together with the WMAP data are represented in the bottom
panels of Fig.~\ref{fig:cls}. The reason for such an important
improvement of the $\chi ^2$ is clear from these figures: the presence
of the oscillations allows a better fit of the cosmic variance
outliers at small scales.
In order to make this statement more quantitative, the difference
between the cumulative $\chi^2$ with and without the oscillations (in
this last case this is nothing but the WMAP cumulative $\chi ^2$) has
been plotted in Fig.~\ref{fig:cumulchi2} as function of the angular
scale $\ell$.
This permits a direct comparison with the Fig.~4 of
Ref.~\cite{spergel}. Clearly, the significant decrease of the $\chi^2$
comes from the outliers around the first acoustic peak which are well
fitted by the oscillating component of our initial power
spectra. Notice also that the outliers at large scales, $\ell<50$, can
not be well fitted in this model due to the strong damping described
above [see Eq.~(\ref{eq:largescalestpl})]. But, as already mentioned
several times, this is not a problem since the outliers that
contributes the most to the $\chi ^2$ are not located at large scales
but around the first acoustic peak.

\par

To conclude this section, let us remark that we have not given
confidence values for the best fit model parameters in
Table~\ref{tbl:bestfit}. This is due to the fact that the
high-accuracy \CAMB version, which is necessary to obtain physical
results when the oscillations are included, takes a few minutes to
compute only one model instead of a few seconds in the standard
inflationary case. Although minima in the $\chi^2$ hypersurface can be
``rapidly'' found with \COSMOMC, the Markov chains take a much longer
time to scan the ``one-sigma'' neighborhood. In addition, as studied
in Ref.~\cite{steen}, the likelihood function looks like a hedgehog in
the parameters space forbidding a smooth convergence of the Markov
chains around a given $\chi^2$ minimum.

\section{Discussion and conclusion}
\label{sect:end}

Given the previous result, $\Delta \chi ^2\simeq15$, the first question
that one may ask is whether the improvement is statistically
significant? We have introduced two new parameters, $\sigma _\zero$
and $x$, and one may wonder whether it was worth it, given the
decrease observed in the $\chi ^2$. Indeed, it could just be due to
statistical fluctuations in the larger parameters space. Since it is
not possible to directly compare the $\chi ^2$ of models with
different degrees of freedom, i.e. with different number of
parameters, we need another reliable statistical test.  The
F--test~\cite{statappl} is an efficient tool to deal with this
problem. This test works as follows. Given a first fit with $n_\one$
degrees of freedom and with a $\chi^2_\one$ and a second fit with
$n_\two<n_\one$ degrees of freedom and a new $\chi ^2_\two<\chi
^2_\one$, the F--test gives the probability that the decrease in the
$\chi ^2$ is only due to statistical fluctuations and not to the fact
that the underlying model is actually a better fit. Generally, one
considers that the improvement is significant if this probability is
less than a few $\%$. For our case, the F--probability is F${}_{\rm
proba}\simeq 0.06\%$.

\par

A possible loophole would be that, since the F--test is only valid for
Gaussian statistics, it could not be applied to the case at hand since
the multipole moments obey a different statistics. However, since our
result comes from relatively small scales, the central limit theorem
applies and the statistics should not deviate too much from the
Gaussian one. Therefore, we expect the F--test to give a rather fair
estimate. Another test has been to check that a spectrum with a
different oscillatory pattern, e.g. with a $k$-dependence instead of
$\ln k$ in the cosine function, does not provides us with a fit as
good as the trans-Planckian one, in the F--test sense and for a same
amount of computed models with \COSMOMC. This is indeed the case since
we have found $\Delta \chi ^2\simeq 4$ only (with two new parameters)
which is not very statistically significant.

\par

Let us now analyze and discuss the result itself in more details. The
first remark is that the baryons contribution is standard since we
have $\OmegaB h^2\simeq 0.025$. It is therefore compatible with Big
Bang Nucleosynthesis even if this is slightly less compatible than the
value obtained without oscillations but this difference does not seem
to be significant. In the same way, the value of $\OmegaCDM$ is not
significantly modified compared to the standard one. The same is true
for the energy scale of inflation, i.e. for $\Ps$. On the other hand,
we see that the value of $\OmegaL$ is particularly high, as the values
of $h$ and the optical depth are. However, this still seems to be
admissible, in particular with the SNIa measurements. However, the
slow-roll parameters differ significantly compared to their standard
values~\cite{peiris,saminf,barger} and, as a consequence, the spectral
index is modified. This is not surprising and should even be expected:
since the trans-Planckian parameters seem to be as statistically
significant as the slow-roll ones, their inclusion in the best fit
search should have an effect on the determination of the slow-roll
parameters. The fact that the best fit cosmological parameters with or
without oscillations are not exactly the same illustrates the fact
that the determination of these cosmological parameters depends on the
shape of the primordial power spectra~\cite{LLMS}.

\par

Let us now study the trans-Planckian parameters in more details. First
of all, we see that the best fit is such that the preferred scale
$\Mc$ is almost four orders of magnitude higher than the Hubble scale
during inflation. However, this does not provide us with a measure of
$\Mc$ since the energy scale of inflation cannot be deduced directly
from the data. Nevertheless, a constraint on $H$ has been given in
Ref.~\cite{saminf} and leads to $\Mc/\mP<0.1$. On the other hand, from
Table~\ref{tbl:bestfit}, one obtains that $4 \times 10^4\sigma
_\zero^2/\sqrt{\epsilon }\simeq 10^{-2}$. This means that the
trans-Planckian amplitude, $|x| \sigma_\zero = 0.29$, is not
compatible with the requirement of negligible backreaction effects
[see Eq.~(\ref{eq:brvalid})]. This is certainly a major difficulty not
for the presence of oscillations in the multipole moments (since we
have proven that this is statistically significant anyway) but rather
for the physical interpretation of those oscillations in terms of
trans-Planckian physics. At this point, two remarks are in
order. First, the previous calculation does not predict what the
modifications coming from the inclusion of the backreaction effects
are and, in particular, it does not tell that the backreaction effects
will modify the power spectra. It just signals when the backreaction
effects must be taken into account. Therefore, despite the fact that
we are not able to prove it, it could very well be that the
backreaction effects do not modify the $k$-dependence of the
oscillations (for instance, it could only renormalize the values of
the trans-Planckian parameters). Clearly, this is a very difficult
technical question since a check of the above speculation would
require a calculation at second order in the framework of the
relativistic theory of cosmological perturbations with the
trans-Planckian effects taken into account. This is beyond the scope
of the present article. Secondly, the physical origin of the
oscillations could be different. For instance, in
Refs.~\cite{BCLH,kaloper}, initial power spectra with oscillations
were also found but with different physical justifications. In the
model of Ref.~\cite{BCLH}, the presence of the oscillations is due to
non-standard initial conditions in the framework of hybrid inflation,
while in Ref.~\cite{kaloper} a sudden transition during inflation is
involved. Of course, it remains to be proven that a different model
could produce an oscillatory pattern similar to the trans-Planckian
one but since, on a purely phenomenological level, it seems that the
significant increase of the likelihood requires superimposed
oscillation in the initial power spectra, this is a question that is
certainly worth studying. Let us also signal that a primordial
bouncing phase, as described for instance in Ref.~\cite{MP}, may also
do the job. Finally, let us remind that the main goal of the paper was
to study whether the idea that oscillations are present in the WMAP
data can be statistically demonstrated and that the trans-Planckian
power spectra used before were just one possible example.

\par

As a conclusion, a word of caution is in order. From the previous
considerations, it is clear that, in the absence of outliers around
the first acoustic peak, there would be no reason, in the F--test
sense, to include oscillations with a large amplitude (i.e. $\vert
x\vert \neq 1$) in the power spectra. Therefore, the future WMAP data
release will be extremely important to check whether the presence of
these outliers is confirmed or if they are just observational
artifacts. However, if the last possibility turns out to be true, it
is clear that it would still be worth looking for small
trans-Planckian effects (i.e. $\vert x\vert =1$) in the future very
high-accuracy Planck data and the ideas put forward in the present
article would still be useful for this purpose.

\vspace{1cm}

\acknowledgments

We wish to thank G.~H\'ebrard, A.~Lecavelier des Etangs, M.~Lemoine,
M.~Peloso, P.~Peter, S.~Prunet and B.~Revenu for helpful comments
and/or careful reading of the manuscript.  We are especially indebted
to S.~Leach for his help and many enlightening discussions. It is also
a pleasure to thank R.~Trotta for his useful advice on the \CAMB and
\COSMOMC armory. We would like to thank the CINES for providing us one
year of CPU--time on their SGI--O3800 supercomputers.

\end{document}